\documentclass[12pt]{article}
\usepackage{amsmath}
\usepackage{graphicx}
\usepackage{feynmf}
\newcommand{\Tr}{\ensuremath{\mathop{\mathrm{Tr}}}}

\newcommand{\del}{\partial}

\begin{document}
\begin{fmffile}{big}

%\begin{titlepage}

%\begin{center}
\title{Free Energy and Phase Transition of the Matrix Model on a Plane-Wave}

\author{Shirin Hadizadeh, Bojan Ramadanovic, \\Gordon W. Semenoff and Donovan Young\\
 Department of Physics and Astronomy, \\University of British
Columbia,\\ 6224 Agricultural Road, \\Vancouver, British Columbia,
Canada V6T 1Z1\\shirin, bramadan, dyoung@physics.ubc.ca~~semenoff@nbi.dk}

\maketitle

\abstract{ It has recently been observed that the weakly coupled
plane wave matrix model has a density of states which grows
exponentially at high energy.  This implies that the model has a
phase transition.  The transition appears to be of first order.
However, its exact nature is sensitive to interactions. In this
paper, we analyze the effect of interactions by computing the
relevant parts of the effective potential for the Polyakov loop
operator in the finite temperature plane-wave matrix model to
three loop order. We show that the phase transition is indeed of
first order. We also compute the correction to the
Hagedorn temperature to order two loops.}

 \vspace{10cm}

%\end{center}
%\end{titlepage}
\newpage
\section{Introduction}

It is conjectured that the full dynamics of M-theory is encoded in
a certain maximally supersymmetric matrix quantum mechanics called
the BFSS matrix model \cite{Banks:1996vh}.   At finite $N$ that
model is thought to describe a discrete light-cone quantization of
M-theory in the infinite momentum frame.  The light-cone is
decompactified by a taking a particular large $N$ limit where the
discrete light-cone momentum $p^+=N/R$ is held constant, with $R$
the radius of the null identification $x^-\sim x^-+2\pi R$.

Recently, a massive deformation of the BFSS matrix model
appropriate to M-theory on a plane wave background has been
studied\cite{Berenstein:2002jq}. The main difference between it
and the BFSS model is their supersymmetry, which in the latter
case is that of a maximally supersymmetric plane wave spacetime
rather than 11-dimensional Minkowski space.  A great advantage of
the plane wave matrix model is that it has a weak coupling regime
where it can be studied systematically using perturbation
theory\cite{Dasgupta:2002hx}-\cite{Klose:2003qc}.  It also has a
powerful supersymmetry algebra which allows the extrapolation of
some perturbative results to the string coupling regime.

The action of the plane wave matrix model is
\begin{eqnarray}
S  =\frac{1}{2R} \int d\tau  \Tr \left(  D {X}^i D {X}^i +
 D X^{\bar a} D X^{\bar a} + i
\psi^{\dagger I \alpha} D \psi_{I \alpha}  + \frac{R^2}{2} [X^i ,
X^j]^2 +\qquad \qquad \right. \nonumber \\ \left. + \frac{R^2}{2}
[X^{\bar a}, X^{\bar b}]^2 + R^2 [ X^{\bar a}, X^i]^2 +
\frac{R^2}{2} [X^i , X^j]^2 + \frac{R^2}{2} [X^{\bar a}, X^{\bar
b}]^2 + R^2 [ X^{\bar a}, X^i]^2 \right. \nonumber \\  \left.
\qquad -R\, \psi^{\dagger I \alpha} \sigma^{\bar a}_\alpha
{}^\beta [X^{\bar a}, \psi_{I \beta}] + \frac{R}{2}
\epsilon_{\alpha \beta} \psi^{\dagger \alpha I} {\sf g}^i_{IJ}
[X^i, \psi^{\dagger \beta J}] - \frac{R}{2} \epsilon^{\alpha
\beta} \psi_{\alpha I} ({\sf g}^{i \dagger})^{IJ} [X^i,
\psi_{\alpha J}] \right. \nonumber \\   \left.
 - \left(\frac{\mu}{3}\right)^2 (X^{\bar a})^2 -
\left(\frac{\mu}{6} \right)^2 (X^i)^2  - \frac{\mu}{4}
\psi^{\dagger I \alpha} \psi_{I \alpha} - i\mu\frac{2R}{3 }
\epsilon_{\bar a \bar b \bar c} X^{\bar a} X^{\bar b} X^{\bar c}
\right)\label{matrixmodelaction}
\end{eqnarray}

\noindent where all of the variables are $N\times N$ matrices. The
indices on the bosonic matrices take values $\bar a,\bar b, \bar
c=1,2,3$, $i,j=4,\ldots,9$.  The U(N) symmetry is gauged.  All
variables transform in the adjoint representation of the gauge
group $X^i\to UX^iU^{\dagger}$, etc. The time derivatives are
covariant, $D = \del_\tau - i\left[A,...\right]$ with an $N\times
N$ Hermitian gauge field $A$. The fermions have $8$ complex
components with $I,J=1,\ldots,4$ and $\alpha,\beta=1,2$. The spin
matrix has the property $ {\sf g}^i ({\sf g}^j)^\dagger + {\sf
g}^j ({\sf g}^i)^\dagger = 2 \delta^{ij} {\bf 1}_{4\times 4} $.
$\epsilon_{\alpha\beta}$ and $\epsilon_{\bar a \bar b \bar c}$ are
antisymmetric tensors.

Note that the last line of (\ref{matrixmodelaction}) contains terms with the
parameter $\mu$ which comes from the plane wave geometry.  That
geometry has metric and constant four-form,
$$ ds^2=-2dx^+dx^-+dx^a
dx^a+dx^idx^i-dx^+dx^+\left(\left(\frac{\mu}{3}\right)^2
(x^a)^2+\left(\frac{\mu}{6}\right)^2 (x^i)^2\right)$$
$$F_{123+}=\mu $$ It is a maximally symmetric solution of
11-dimensional supergravity for any value of the parameter $\mu$.
If we set $\mu$ to zero we recover the Minkowski metric.

The plane wave matrix model (\ref{matrixmodelaction}) can be systematically
analyzed in perturbation theory.     When $N$ is large, the
expansion parameter is \cite{'tHooft:1973jz}
\begin{equation}
\lambda = \left( \frac{3R}{\mu}\right)^3 N
\label{lambda}\end{equation}   Note that all dimensional
quantities are in units of the eleven dimensional Planck length
which we have set to one.

Aside from being a model of M-theory on a plane-wave background,
there is an interesting connection between the matrix model
(\ref{matrixmodelaction}) and four dimensional ${\cal N}=4$
supersymmetric Yang-Mills theory.  At the classical level, the
degrees freedom and dynamics described by (\ref{matrixmodelaction})
are a consistent truncation of the Yang-Mills theory which keeps
only those modes which are invariant under a certain $SU(2)$
subalgebra of the full $SU(2,2|4)$ superconformal algebra.
In the planar limit, this truncation seems to also hold at the
one-loop level~ \cite{Kim:2003rz}. It is then
natural to speculate about whether it would inherit some of the
properties of ${\cal N}=4$ supersymmetric Yang-Mills theory, such
as integrability of the large $N$ limit, at least in the leading
orders of perturbation theory. This has been studied in detail in
recent work \cite{Klose:2003qc}.  It has been shown that the form
of the dilatation operator for supersymmetric Yang-Mills theory
which has been conjectured using integrability as an input, when
restricted to the appropriate sector of the theory, is identical
to the Hamiltonian of the plane wave matrix model, at least for
computing energy levels to three loop order.

\subsection{Thermodynamics}

Some aspects of the thermodynamics of the plane wave matrix model
have been studied before
\cite{Furuuchi:2003sy}-\cite{Spradlin:2004sx}. It was shown in
\cite{Furuuchi:2003sy} that, at weak coupling, and in the large N
limit, there is a deconfinement phase transition at a particular
temperature. The use of the word ``deconfinement'' in a theory
where there is no spatial extent over which particles can be
separated must be justified carefully. It is possible to
distinguish the two phases by the  behavior of the free energy,
$F[T]$.  In a weakly coupled theory, the free energy should be
proportional to the number of degrees of freedom.  In the
confining phase the number of degrees of freedom at a given energy
are of order the number of color singlets, which in the lower
energy parts of the spectrum does not grow with the rank of the
gauge group, $N$, but is of order one.
 In a deconfined phase, the number of degrees of freedom
is the number of elements of the matrices, which is of order
$N^2$. Thus, we would expect
\begin{eqnarray} \lim_{N\to\infty}~~  \frac{F}{N^2}&=&0 ~{\rm confined}
\nonumber \\ \lim_{N\to\infty}~~  \frac{F}{N^2}&\neq & 0 {\rm~
deconfined}\
 \label{critereon1}\end{eqnarray}
When interactions are turned on, the large $N$ limit which we are
using is the 't Hooft limit \cite{'tHooft:1973jz}, where the
coupling constants are also scaled as $N$ is taken large so that
$\lambda$ defined in (\ref{lambda}) is held constant. Note that
this limit is different from the large $N$ limit with $N/R$ held
constant which would decompactify the null direction in M-theory.

In a gauge theory such as (\ref{matrixmodelaction}), where all
variables transform in the adjoint representation, there is an
order parameter for confinement called  the Polyakov
loop~\cite{Polyakov:vu,Susskind:1979up}. It is the trace of the
holonomy of the gauge field around the finite temperature
Euclidean time circle,
\begin{equation}
P=\frac{1}{N}{\rm Tr}\left( e^{i\oint A} \right)
\label{ploop}\end{equation}  It is related to the difference of
free energies  of the state where an additional fundamental
representation quark is introduced and the state without the
quark,
$$ F_q[T]-F_0[T]= -T~\ln\left<P\right>$$  The free energy of a free
quark in the confined phase should be large compared to that in
the deconfined phase.  In fact,
 \begin{equation}
\begin{matrix} \left< P\right>~~=0 &~{\rm confined}
\cr \left< P\right>~~\neq 0 &{\rm~ deconfined}\cr
\end{matrix} \label{critereon2}\end{equation}  is
interpreted as requiring infinite energy to insert a fundamental
representation quark into the system when it is in the confining
phase, whereas it is finite in the deconfined phase. The Polyakov
loop operator is widely used to study finite temperature
de-confining phase transitions in higher dimensional gauge
theories, \cite{McLerran:1980pk}-\cite{Dumitru:2003hp}

A phase transition where both criteria for deconfinement
(\ref{critereon1}) and (\ref{critereon2}) occur has been found in
the large $N$ limit of the matrix model \cite{Furuuchi:2003sy}. It
has been argued to be generic to the large $N$ limit of matrix
quantum mechanics when the matrices have a gapped spectrum.  The
transition temperature is of order of the mass gap.

An intuition for why this phase transition occurs can be gained
from counting matrix degrees of freedom~\cite{Furuuchi:2003sy}. As
a simple example, consider a set of $d$ matrix harmonic
oscillators with Hamiltonian
\begin{equation}\label{Hamiltonian}H=\sum_{i=1}^d\omega{\rm Tr}
(\alpha_i^{\dagger}\alpha_i)\end{equation} and matrix-valued
creation and annihilation operators with algebra
\begin{equation}\label{comm} \left[
\alpha^i_{ab},\alpha^{j\dagger}_{cd}\right]=\delta^{ij}\delta_{ad}
\delta_{bc}\end{equation} States are created  the creation
operators  $\alpha^{i\dagger}_{ab}$ operating on a vacuum $|0>$.
To get a state with energy $E=n\omega$ we must act with $n$
creation operators.

The analog of gauge symmetry is to require a physical state
condition of invariance under the unitary transformation
$$\alpha^{\dagger}\to u\alpha^{\dagger}u^{\dagger}~~,~~
\alpha\to u\alpha u^{\dagger}$$ where $u\in U(N)$. We assume that
the vacuum state is invariant.  Then, physical states are created
by operating with invariant combinations of creation operators. In
the limit $N\to\infty$ all such combinations are traces

\begin{equation} \left[{\rm
Tr}\left(\alpha_i^{\dagger}\right)\right]^{n_1}\left[{\rm
Tr}\left(\alpha_{j_1}^{\dagger}
\alpha_{j_2}^{\dagger}\right)\right]^{n_2}\left[{\rm Tr}\left(
\alpha_{k_1}^{\dagger}\alpha_{k_2}^{\dagger}
\alpha_{k_3}^{\dagger}\right)\right]^{n_3}\ldots|0>\label{state}\end{equation}
where the energy is $$E=\omega(n_1+2n_2+3n_3+\ldots)$$ The number
of these traces with a fixed energy, $E$, does not scale like
$N^2$ as $N\to\infty$, instead it approaches a constant as $N$ is
taken large.  Thus at low enough temperatures, the free energy
should approach an $N$-independent constant as $N$ is taken large.

However, the number of independent traces does increase rapidly
with the energy $E$. It has been shown~\cite{Furuuchi:2003sy}
that, in the large $N$ limit, the oscillator has a Hagedorn
density of states at high energy, $$\rho(E)\sim \frac{1}{E}~e^{
E/T_H}$$ where the Hagedorn temperature is
\begin{equation}\label{hagtemp}T_H=\omega/\ln d\end{equation}
This mirrors a similar
discussion for weakly coupled Yang-Mills
theory~\cite{Sundborg:1999ue}-\cite{Spradlin:2004pp}.

At temperatures higher than $T_H$, the thermodynamic canonical
ensemble does not exist. It could be made to exist by keeping $N$
large but finite. That would cut off the exponential growth in the
asymptotic density of states at some large energy, of order $N^2$.
Then we could consider a temperature that is greater than $T_H$.
Both the energy and entropy would be dominated by states at and
above the cutoff scale and would be of order $N^2$. The divergence
of the free energy $\sim N^2$ occurs as we take the limit
$N\to\infty$ at constant temperature (noting that the Hagedorn
temperature does not depend on $N$).

It was shown in ref.~~\cite{Furuuchi:2003sy} that, from a
different point of view, this behavior of the matrix model can
also be found as a large $N$ Gross-Witten  type of phase
transition~\cite{Gross:1980he,Semenoff:1996vm} which is familiar
in unitary matrix models. In the next Section, we will discuss
this in more detail.

 \section{One loop}

 Now, let us consider a perturbative expansion of the plane wave
matrix model at finite temperature.  The partition function has
Euclidean path integral representation similar to those of finite
temperature quantum field theories.  We consider the theory in
Euclidean space where time is compact and identified as $$
\tau\sim \tau+\beta$$ with $$\beta=1/T$$  $T$ is the temperature.
The partition function is given by the functional integral
\begin{equation} Z=\int [dA][dX^i][d\psi]e^{-\int_0^\beta d\tau
L[A,X^i,\psi]}\end{equation} where $L$ is the Euclidean time
Lagrangian
\begin{equation}
\begin{split}
L &=\frac{1}{2R} \Tr \left( D {X}^i D {X}^i +
 D X^{\bar a} D X^{\bar a} -
\psi^{\dagger I \alpha} D \psi_{I \alpha}
\right)\\
 & \!   + \frac{1}{ 2R}  \Tr \left(
 \left(\frac{\mu}{3}\right)^2 (X^{\bar a})^2
+ \left(\frac{\mu}{6} \right)^2 (X^i)^2  + \frac{\mu}{4}
\psi^{\dagger I \alpha} \psi_{I \alpha} + i\mu\frac{2R}{3}
\epsilon_{\bar a \bar b \bar c} X^{\bar a} X^{\bar b} X^{\bar c}
\right. \\
& \qquad + R\,\psi^{\dagger I \alpha} \sigma^{\bar a}_\alpha
{}^\beta [X^{\bar a}, \psi_{I \beta}] - \frac{R}{2}
\epsilon_{\alpha \beta} \psi^{\dagger \alpha I} {\sf g}^i_{IJ}
[X^i, \psi^{\dagger \beta J}] + \frac{R}{2} \epsilon^{\alpha
\beta}
\psi_{\alpha I} ({\sf g}^{i \dagger})^{IJ} [X^i, \psi_{\alpha J}] \\
& \left. \qquad -  \frac{R^2}{2} [X^i , X^j]^2 - \frac{R^2}{2}
[X^{\bar a}, X^{\bar b}]^2 - R^2 [ X^{\bar a}, X^i]^2 \right)
\end{split}
\end{equation}
  The bosonic and fermionic variables have periodic
and antiperiodic boundary conditions, respectively
$$A(\tau+\beta)=A(\tau) ~~,~~X^i(\tau+\beta)=X^i(\tau)
~~,~~\psi(\tau+\beta)=-\psi(\tau) $$ Since the boundary conditions
for fermions and bosons are different, supersymmetry is broken
explicitly. Of course this is expected at finite temperature where
bosons and fermions have different thermal distributions.
Supersymmetry is restored in the zero temperature limit.  We will
see the results of this explicitly in the following. A parallel
discussion of the BFSS matrix model at finite temperature can be
found in ref. \cite{Ambjorn:1998zt}-\cite{Makeenko:1999hp}.

\subsection{Gauge fixing}

To begin, we must fix the gauge.  It is most convenient to use the
gauge freedom to make the variable $A$ static and diagonal,
$$
\frac{d}{d\tau}A_{ab}=0 ~~,~~ A_{ab}=A_a\delta_{ab}
$$
Once this is done, the remaining degrees of freedom of $A$ are the
time-independent diagonal components, $A_a$. We shall see that
they eventually appear in the form $\exp\left( i\beta A_a\right)$.

The Faddeev-Popov determinant for the first of these gauge fixings
is\footnote{Using zeta-function regularization,
$${\det}'\left(-\frac{d}{d\tau}\right)=\beta$$.}
\begin{equation}\label{fp1}
{\det}'\left(-
\frac{d}{d\tau}\left(-\frac{d}{d\tau}+i(A_a-A_b)\right)\right) =
{\det}'\left(-
\frac{d}{d\tau}\right){\det}'\left(-\frac{d}{d\tau}+i(A_a-A_b)\right)
\end{equation}
where the boundary conditions are periodic with period $\beta$.
The prime means that the zero mode of time derivative operating on
periodic functions is omitted from the determinant. Once the gauge
field is time-independent, we do the further gauge fixing which
makes it diagonal. The Faddeev-Popov determinant for diagonalizing
it is the familiar Vandermonde determinant,
$$
\prod_{a\neq b}|A_a-A_b|
$$
This is also just the factor that the time independent zero mode
would contribute to the second of the determinants in (\ref{fp1}).
Including it gives the determinant
\begin{equation}\label{fp2}
\prod_{a\neq b} {\det}'\left(-
\frac{d}{d\tau}\right)\det\left(-\frac{d}{d\tau}+i(A_a-A_b)\right)
\end{equation}
where there is now no prime on the second factor.  These
determinants can be found explicitly. We will do this shortly.

\subsection{Classical ground states}

We will perform a semiclassical expansion of the free energy in
the sector of the theory with the classical ground state
$X^a=0=X^i$ in the large $N$, 't Hooft limit. This is a double
expansion.  First, it keeps only planar Feynman diagrams, which is
the leading order in an expansion in $1/N^2$. Secondly, it is
perturbative in that it keeps those diagrams which are of low
orders in the coupling constant. Since $N$ is large, the
appropriate coupling constant is $\lambda$ defined in
eqn.~(\ref{lambda}).

The configuration $X^a=0=X^i$ has zero classical energy. In fact,
because of supersymmetry, the ground state energy of the theory
quantized beginning with this vacuum is zero to all orders in
perturbation theory. This will provide us with an important check
of our finite temperature computations.   Of course, temperature
breaks supersymmetry and the free energy of the thermodynamic
state, which is what we compute, is non-zero. However, we will
always be able to check whether the zero temperature limit of the
free energy, which is the ground state energy, vanishes.  We find
that its expansion in the coupling constant $\lambda$ indeed does
so to order $\lambda^2$.

In this subsection, for completeness, we comment on the fact that
there are a large number ($\to \infty$ as $N\to \infty$) of other
ground states with zero energy. To see this, observe that the
classical potential in (\ref{matrixmodelaction}) can be written in
the form
\begin{eqnarray}
V=\frac{R}{2}{\rm Tr}\left[ \left(
\frac{\mu}{3R}X^a+i\epsilon^{abc}X^bX^c\right)^2
+\frac{1}{2}\left( i[X^{i},X^{j}]\right)^2 +
\right.\nonumber\\
\left. +\left(i[X^{i},X^a]\right)^2+
\left(\frac{\mu}{6R}\right)^2(X^{i})^2 \right]
\end{eqnarray}
It has isolated global classical minima ($V=0$) where
\begin{equation}
X^{i}_{\rm cl}=0 ~~,~~ X^a_{\rm cl}= \frac{\mu}{3R}J^a
\label{fs}\end{equation} and $J^a$ form  an N-dimensional
representation of the SU(2) algebra,
$[J^a,J^b]=i\epsilon^{abc}J^c$. In addition, the classical
solution for gauge field must obey the equation
$$
\left[ A_{\rm cl} , J^a_{\rm cl} \right] = 0
$$

If $J^a$ are an irreducible representation of SU(2), by Schur's
Lemma, $A_{\rm cl}$ must be proportional to the unit matrix
$A_{\rm cl}=c\cdot{\cal I}$ and a symmetry of the theory allows us
to set $c=0$. The gauge symmetry is realized by the Higgs
mechanism and fluctuations of the gauge field are all massive. On
the other hand, when the representation is reducible, there are
gauge fields which commute with the condensate. The parts of
$A_{\rm cl}$ which commute with the condensate remain undetermined
by the classical equations. The volume of the space of all
possible such $A_{\rm cl}$ forms a moduli space of the classical
solutions which must still be integrated over, even to obtain the
leading order in the semi-classical approximation to the partition
function.

The implications of the solutions (\ref{fs}) have been discussed
in detail in ref.~\cite{Maldacena:2002rb}.  They were interpreted
in terms of the spherical membrane and transverse spherical
5-branes which exist on the 11-dimensional plane wave background.

In refs.~\cite{Furuuchi:2003sy,Semenoff:2004bs} it was  argued
that the phase transition that we shall study here occurs only
when there is a residual gauge symmetry and only in the limit
where the rank of the residual gauge group goes to infinity.
According to ref.~\cite{Maldacena:2002rb}, this is the limit of
the theory which describes 5-branes.  Perturbation theory in that
limit is governed by a 't Hooft coupling similar to (\ref{lambda})
with $N$ replaced by the rank of the residual gauge group.
Ref.\cite{Maldacena:2002rb} argued that, in the 't Hooft limit,
the barrier between the degenerate vacua becomes infinitely high.
They also argued that the limit decouples the 5-brane from other
degrees of freedom and focuses on its internal dynamics.

The results of refs.~\cite{Furuuchi:2003sy,Semenoff:2004bs} can be
interpreted as saying that the phase transition of the matrix
model occurs only in 5-brane states and not in membrane states,
and it seems to be associated with internal dynamics of the
5-branes.  Of course, refs.~\cite{Furuuchi:2003sy,Semenoff:2004bs}
analyzed only the weak coupling limit of the matrix model which is
far from the limits which are conjectured to describe supergravity
of an 11-dimensional spacetime continuum where the 5-brane would
live. In order to apply it directly to any known behavior of
5-branes, it would have to be extrapolated to strong 't Hooft
coupling. This is of course a difficult problem. The perturbative
expansion in the present paper perhaps gives an indication that
the phase transition is of first order but much more would have to
be done to answer the question of persistence or nature of the
phase transition in the 5-brane regime.  One interesting extension
of the present work would be to examine the dependence of the
nature of the phase transition on the number of 5-branes.  For $k$
coincident 5-branes, $X^a$ contains $N/k$ $k$-dimensional
representations of SU(2), with k being held fixed in the large N
limit.  $k$-dependence of the phase transition temperature was
investigated in ref.~\cite{Furuuchi:2003sy}.

\subsection{Semiclassical expansion}

If we expand about the classical vacuum  $X_{\rm cl}^a=0=X_{\rm
cl}^i$, we find the partition function in the 1-loop approximation
is
\begin{equation}
Z= \int dA_a\prod_{a\neq b} \frac{ {\det}'\left(
-d/d\tau\right){\det}\left(-D_{ab}\right)
 \det^8\left(
-D_{ab}+\frac{\mu}{4}\right) }{ \det^{3/2}\left(
-D_{ab}^2+\frac{\mu^2}{9}\right) \det^{3} \left(
-D_{ab}^2+\frac{\mu^2}{36}\right) }
\end{equation}
where $D_{ab}=\frac{d}{d\tau}-i(A_a-A_b)$. The first two terms in
the numerator are the Faddeev-Popov determinant.  The third term
comes from fermions whereas the denominator is from bosons.
Using the formula
$$
\det\left(
-\frac{d}{d\tau}+\omega\right)=2\sinh\frac{\beta\omega}{2}
$$
with periodic boundary conditions and
$$
\det \left(
-\frac{d}{d\tau}+\omega\right)=2\cosh\frac{\beta\omega}{2}
$$
with antiperiodic boundary conditions, we can
write\footnote{Because the matrix model action
(\ref{matrixmodelaction}) is invariant under replacing $A$ by $A$
plus a constant times the unit matrix, we see that the integrand
in eqn.(\ref{almostpartf}) is indeed invariant under translating
all values of $A_a$ by the same constant.}
\begin{equation}\label{almostpartf}
Z= \int_{-\pi}^{\pi} \prod_{a=1}^N \frac{d\left( \beta
A_a\right)}{2\pi} \prod_{a\neq b} \frac{
[1-e^{i\beta(A_a-A_b)}][1+e^{-\beta\mu/4+i\beta(A_a-A_b)}]^8
 }{ [1-e^{-\beta\mu/3+i\beta(A_a-A_b)}
 ]^3[1-e^{-\beta\mu/6+i\beta(A_a-A_b)}]^6  }
\end{equation}

 Note that, because of supersymmetry, the zero temperature
 ($\beta\to\infty$) limit of the partition function is one.
 It also has a symmetry under replacing $e^{-\beta\mu}$ by
 $1/e^{-\beta\mu}$.

We must now do the remaining integral when $N\to\infty$. There are
$N$ integration variables $A_a$ and the action, which is the
logarithm of the integrand is generically of order $N^2$ which is
large in the large $N$ limit. For this reason, the integral can be
done by saddle point integration. This amounts to finding the
configuration of the variables $A_a$ which minimize the effective
action:
\begin{eqnarray}
S_{\rm eff}=\sum_{a\neq b} \left( -\ln
[1-e^{i\beta(A_a-A_b)}]-8\ln[1+e^{-\beta\mu/4+i\beta(A_a-A_b)}]
 +\right.\nonumber\\ \left.~~~~+3\ln[1-e^{-\beta\mu/3+i\beta(A_a-A_b)}
 ]+6\ln[1-e^{-\beta\mu/6+i\beta(A_a-A_b)}]\right)
 \label{effectiveaction}
\end{eqnarray}
To study the minima, it is illuminating to Taylor expand the
logarithms in the phases (this requires some assumptions of
convergence for the first log)
\begin{eqnarray}
S_{\rm eff}= \sum_{n=1}^\infty \frac{
1-8(-)^{n+1}r^{3n}-3r^{4n}-6r^{2n}}{n}~\phi_{-n}\phi_n \label{eff}
\end{eqnarray}
Here, $$ r=\exp\left( -\beta\mu/12\right) $$ and
\begin{equation}\label{momenta}
\phi_n~=~\frac{1}{N}\sum_{a=1}^N e^{in\beta A_a} \end{equation}
Recalling (\ref{ploop}), we note that $\phi_n$ are multiply wound
Polyakov loop operators evaluated in the static, diagonal gauge.
The zeroth moment is normalized
\begin{equation}\label{constraints1}
\phi_0=1
\end{equation}
The other elements are constrained by sum rules.  The density
defined by
\begin{eqnarray} \rho(\chi)&=&\frac{1}{N}\sum_{a=1}^\infty
\delta(\chi-\beta A_a)~\geq 0 \nonumber \\&=&\sum_n e^{-2\pi i
n\chi}\phi_n \label{constraints2}\end{eqnarray} is a non-negative
function. For example, if only $\phi_0$ and $\phi_{\pm 1}$ are
nonzero, (\ref{constraints2}) implies that $|\phi_1|\leq 1/2$.

In this one-loop approximation, the action is quadratic in the
Polyakov loops. When all coefficients of the quadratic terms are
positive, the action is minimized by $\phi_n=0$ for $n\neq 0$.
This is the confining phase.  When a coefficient becomes negative,
the effective action is minimized with one of the loops nonzero.
The result is a condensation of the loops.

As we raise the temperature from zero (and lower $\beta$ from
infinity), the first mode to condense is $n=1$. This occurs when
$$r_C=1/3~~\to ~~ T_C=\frac{\mu}{12\ln3}\approx .0758533 \mu$$
and $\phi_1\neq 0$ when $T>T_C$. There are some speculations about
the nature of the distribution of the angles $\beta A_a$ in the
deconfined phase in ref.~\cite{Semenoff:2004bs}.

 Note that this condensation breaks a U(1) symmetry. This is associated with
the center of the gauge group $U(1)\in U(N)$.  It arises from the
fact that all variables are in the adjoint representation. In the
Euclidean path integral,   gauge transformations $X(\tau)\to
U(\tau)X(\tau)U^{\dagger}(\tau)$ must preserve the periodicity of
the dynamical variables.  They therefore must be periodic up to an
element of the center, $U(\beta)=e^{i\theta}U(0)$. The Polyakov
loop, on the other hand, being the holonomy on the time circle,
does transform as $P\to e^{i\theta}P$.

Even once the static, diagonal gauge is fixed, there is a vestige
of this symmetry where $\beta A_a\to \beta A_a+\theta$ or
$\phi_n\to e^{in\theta}\phi_n$. This symmetry restricts the form
of the effective action for Polyakov loops, so that the term with
$\phi_{k_1}\ldots\phi_{k_n}$ must have $\sum{k_i}=0$. It is a good
symmetry of the confined phase and it is spontaneously broken in
the deconfined phase.  The Polyakov loop operator is an order
parameter for this symmetry breaking.

\section{Higher loop order}

In the previous Section, we have  computed the effective action as
a function of  the variables $\phi_n$ in the one-loop
approximation. The result is the quadratic potential in
eqn.~(\ref{eff}).  To find the stable phase, it is necessary to
find those values of $\phi_n$ which minimize (\ref{eff}), subject
to the constraint (\ref{constraints2}).

The phase transition occurs when the curvature of the Gaussian
potential for $\phi_1$ develops a vanishing moment.  This is a
first order phase transition \cite{Semenoff:1996vm}. Unlike
generic first order phase transitions, it occurs at the point
where the curvature of the potential first goes to zero. As a
result, there is no energy barrier separating the two phases and
no coexistence region of the ordered and disordered phases.

These features of the phase transition are very sensitive to
higher order corrections. In the following, we shall take into
account the leading effect of higher order corrections.

Since, as the temperature is raised from zero,  the variable
$\phi_1$ is the first to condense, we focus on its effective
action.
 We have computed the relevant parts of the effective
action up to three loop order.

\noindent First of all, the $\phi_n$-independent part of the effective action (what one
obtains by putting $\phi_n=0$ for $n\neq 0$) vanishes up to
order $\lambda^2$.

\noindent The relevant parts which depend on $\phi_n$ have the form

\begin{equation}
\frac{1}{N^2}S_{\text{eff}} =   \Delta_1 (r) |\phi_1|^2
+\Delta_2(r)|\phi_2|^2+ ~\lambda ~P_1(r) \, \biggl( \phi_1 \phi_1
\phi_{-2} + \text{c.c.} \biggl) + ~\lambda^2~ P_2(r) \, |\phi_1|^4
+ \ldots \label{approxaction}\end{equation}

\noindent The coefficients in  (\ref{approxaction}) are

\begin{eqnarray}
\Delta_1(r) =\left[1-8r^3-3r^4-6r^2\right] - 24\lambda~\left[
\ln(r) \, r^2 (r^2+1)
(r+1)^4 \right]- ~~~~~~~~~~~~~~\nonumber \\
-3\lambda^2r^2\,\left[ ~ \ln(r)^2~ \left( 68\, r^{10}+352\,
r^9+904\, r^8+1536\, r^7+2256\, r^6 +~~~~~\right. \right.
\nonumber
\\ \left. \left. +3104\, r^5+ 4120\, r^4+2304\, r^3+928\,
r^2+192\, r+16\right) -~~~\right. \nonumber
\\
-\left.   \ln(r)~\left( 27\, r^{10}+152\, r^9+390\, r^8+640\, r^7+
915\, r^6+1232\, r^5 +\right.\right. \nonumber \\ \left. \left.
+1748\, r^4+1184\, r^3+466\, r^2+440\, r+102 \right)~ \right]+...
\end{eqnarray}
\begin{equation}
\Delta_2(r) = \frac{1}{2} \left(1+  8 r^6 - 3 r^8 - 6
r^4\right)+\ldots~~~~~~~~~~~~~~~~~~~~~~~~~~~~~~~~~~~~~~~~~~~~~
\end{equation}

\begin{equation}
P_1(r) =-12 \ln(r) \, r^4 (2 r^4-4 r^3+3 r^2-4 r+5) (r+1)^4
\end{equation}

\begin{eqnarray}
 P_2(r)  =3 r^4 \left[~-\ln(r)^2~\left(~136\,
r^{12}+512\, r^{11}+704\, r^{10} -1308\, r^8-1376\, r^7+1560\, r^6
\right. \right. \nonumber \\ \left. \left. +6400 \, r^5+10896\,
r^4+8096\, r^3+2136\, r^2+1536\, r+240~\right) \right. + \nonumber
\\ \left.
+ \ln(r)~\left(~27\, r^{12} + 120\, r^{11} + 166\, r^{10} - 32\,
r^9 - 271\, r^8 - 16\, r^7 + 1044\, r^6 + \right. \right. \nonumber \\
\left. \left. + 2624\, r^5 + 4036\, r^4 + 3256\, r^3 + 774\, r^2 +
768\, r  + 944~\right)~ \right]+\ldots
\end{eqnarray}

\noindent Eliminating $\phi_2$ using its equation of motion, we
obtain the effective action for $\phi_1$, in the large $N$ limit,
and to order $\lambda^2$:

\begin{equation}\label{final}\frac{1}{N^2}
S_{\text{eff}} =   \Delta_1 (r)|\phi_1|^2 + \lambda^2\left( P_2(r)
- \frac{ \left[ P_1(r)  \right]^2 } { \Delta_2(r) } \right) \,
|\phi_1|^4 + \ldots
\end{equation}

\subsection{Phase transition}

As we raise the temperature from zero, the quadratic term in
$\phi_1$ in the effective action vanishes at the critical value of
$r$,

\begin{equation}
r_c = \frac{1}{3} + \lambda \frac{2^6 \cdot 5}{3^5} \ln(3)
- \lambda^2 \left[ \frac{11\cdot 53 \cdot 3061}{2^2\cdot 3^9} \ln(3)
                 + \frac{13^2\cdot 1867}{2^4\cdot 3^8} \ln(3)^2 \right]
+ \ldots
\end{equation}
which translates to the critical temperature
\begin{equation}\label{tc}
T_c = \frac{\mu}{12\ln(3)} \left[ 1 + \lambda\frac{2^6\cdot 5}{3^4}
-\lambda^2 \left( \frac{23\cdot 19927}{2^2\cdot 3^7}
+ \frac{1765769}{2^4 \cdot 3^8}\ln(3) \right) + \ldots \right]
\end{equation}

The zeroth order term in the critical temperature is the
one found in \cite{Furuuchi:2003sy}.  The term of first order in
$\lambda$ agrees with the result quoted in ref.~\cite{Spradlin:2004sx}.

Also, from (\ref{final}) we see that the quartic term in $\phi_1$
is negative over the entire range $0<r<1$.  This means that the
phase transition is of first order.  When $r$ is just less than
the critical $r_C=1/3$, the extremum of the effective action at
$\phi_1=0$ is only a local minimum.   The effective action has a
second zero when $$\vert\phi_1\vert^2=-\frac{1}{\lambda^2}\frac{
\Delta_1(r)}{P_2(r)-P_1^2(r)/\Delta_2(r)}$$  Higher order terms in
the effective action are individually small at this value of
$|\phi_1|^2$ when $-\frac{
\Delta_1(r)}{P_2(r)-P_1^2(r)/\Delta_2(r)}<<1$.  We are further
constrained by the fact that $\vert\phi_1\vert\leq 1$.  This requires
that $-\frac{
\Delta_1(r)}{P_2(r)-P_1^2(r)/\Delta_2(r)}<\lambda^2<<1$.  The number $-\frac{
\Delta_1(r)}{P_2(r)-P_1^2(r)/\Delta_2(r)}$ is
less than $0.10$ in the range $0.2555<r\leq 1/3$ and is less than 0.001 in
the range $0.3174<r\leq 1/3$.

If $r$ is sufficiently close to $r_c$,
we can reliably say that the absolute minimum of the potential is
not at $\phi_1=0$ but is elsewhere.  This sets an upper bound on
the transition temperature $$T_{\rm crit.}
 <T_c=\frac{\mu}{12\ln(3)}$$
The tunnelling barrier for bubble nucleation during the first order
phase transition is of order $1/\lambda^2$.

\section{Conclusions}

We have found that the phase transition in the weakly coupled
plane wave matrix model is indeed of first order.
As the temperature is raised from zero, the curvature contained in
the quadratic term in the effective action still vanishes at some
critical temperature.   However, before that point is reached,
when there is still an energy barrier between the two phases, the
deconfined phase becomes the lower energy state.  This is the
generic behavior at a first order phase transition.  In fact, this
behavior is seen in other adjoint matrix models
\cite{Semenoff:1996xg}-\cite{Gattringer:1996fi}.  It is also the
behavior that is seen in the collapse of Anti de Sitter space to a
black hole, which is thought to be the analog of this phase
transition in supergravity of a similar deconfinement in ${\cal
N}=4$ supersymmetric Yang-Mills theory \cite{Witten:1998zw}.

Our analysis does not allow us to compute the first order phase transition
temperature accurately, only to deduce that it is of first order.
It does, however, allow us to compute corrections to the Hagedorn
temperature.   This is the temperature at which, if the confining
phase is superheated beyond where it is a global minimum of the
free energy, it eventually becomes perturbatively unstable.  It is
just the place where the corrected curvature of the effective
action vanishes.

To conclude, we summarize our results:
\begin{itemize}
\item{}We have computed the full free energy to two loop order:

\begin{equation*}
\begin{split}
&S_\text{eff}^\text{2-loops} = -\frac{27\beta g^{2}}{4\mu^{2}}
\sum_{abc}\Biggl( -\frac{(1-r^8)}{C_{ab}^{\omega_{1}}C_{ca}^{\omega_{1}}}
-20\frac{(1-r^4)}{C_{ab}^{\omega_{2}}C_{ca}^{\omega_{2}}}
-12\frac{(1-r^8)(1-r^4)}{C_{ab}^{\omega_{1}}C_{ca}^{\omega_{2}}}\\
+&\frac{(r^8+4r^4+1)(r^4-1)^4 + \left[\cos\beta A_{a b}+\cos\beta
A_{b c}+\cos\beta A_{ca}\right]
2r^4(r^4-1)^4}{C_{ab}^{\omega_{1}}C_{bc}^{\omega_{1}}C_{ca}^{\omega_{1}}}\\
+&16 \frac{r^3(r^4-r^2+1)(r^4-1)^2(r^2+1) \left[\cos\beta A_{a
b}+\cos\beta  A_{b c}\right] + r^6(r^4-1)^2\left[2+2 \cos\beta
A_{ca}\right] }
{\bar C_{ab}\bar C_{bc}C_{ca}^{\omega_{1}}}\\
+&32\frac{r^3(r^4-1)^2(r^2+1)\left[\cos\beta A_{a b}+\cos\beta
A_{b c}\right] +r^2(r^8-1)(r^4-1)\cos\beta A_{ca} +
(r^4-1)^2(r^8+1) } {\bar C_{ab}\bar C_{bc}C_{ca}^{\omega_{2}}}
\Biggr)
\end{split}
\end{equation*}

Where we have:

\begin{equation}
\begin{split}
C_{ab}^{\omega_1} &= 1 - 2\,r^4\,\cos{\beta A_{ab}} + r^8\\
C_{ab}^{\omega_2} &= 1 - 2\,r^2\,\cos{\beta A_{ab}} + r^4\\
{\bar C}_{ab} &= 1 + 2\,r^3\,\cos{\beta A_{ab}} + r^6
\end{split}
\end{equation}

This expression can be restated to exhibit its $\phi_n$
dependence:

\begin{equation}
\begin{split}
\frac{1}{N^2}\,S_{\text{eff}}^{\text{2-loops}} 
&= 3\lambda \ln(r) \, \sum_{mn} \phi_n \, \phi_m \, \phi_{-m-n} \Biggl[
-4 \frac{ r^8-r^4+1 }{ (r^4+1)^2 } r^{4|n|+4|m|}\\
&+16 \frac{ (r^4-r^2+1)(r^2+1)^2 }{ (r^4+r^2+1)(r^4+1) } \,(-1)^n \, r^{3|n|+4|m|} 
-16 \frac{ (r^2+1)^2 }{ r^4+r^2+1 } \, (-1)^{n+m}\, r^{3|n|+3|m|} \\
&+32 \frac{ (r^2+1)^2 }{ r^4+r^2+1 } \, (-1)^n \,r^{3|n|+2|m|}
-12 \, r^{4|n|+2|m|} -20\, r^{2|n|+2|m|}\\
&-4\frac{ (r^4-1)(r^8+r^4+1) }{ (r^4+1)^3 } \, F_{mn}(4,4)
+16 \frac{ (r^2+1)(r^{10}-1) }{ (r^4+1)(r^4+r^2+1)^2 } \, (-1)^m\,F_{mn}(3,4)\\
&-16 \frac{(r^2+1)(r^6-r^4+r^2-1) }{ (r^4+r^2+1)^2 } \, (-1)^m\,F_{mn}(3,2)
\Biggr]
\end{split}
\end{equation}

\noindent Where we define the function $F_{mn}(a,b)$ in the following manner:

\begin{equation}
F_{mn}(a,b) = \left\{ \begin{aligned}
F^1_{mn}(a,b) &\qquad m,n \geq 0 \quad \text{or} \quad m,n < 0\\
F^2_{mn}(a,b) &\qquad n < 0, m \geq -n\quad \text{or} \quad n \geq 0, m < -n\\
F^3_{mn}(a,b) &\qquad m < 0, m \geq -n \quad \text{or} \quad m \geq 0, m < -n
\end{aligned}
\right.
\end{equation}

\noindent And where we have:

\begin{equation}
\begin{split}
F^1_{mn}(a,b) = r^{a(2+n+m) + b} &\Biggl[
\frac{ r^{b(n+m)-an} }{ 1-r^{2a+b} } + \frac{ r^{-b-2a+an} }{ 1-r^{2a+b} }
+\frac{ r^{-2a-an+b(n+m)} }{ r^b-1 } \\
&- \frac{ r^{-2a-an+bn} }{ r^b-1 } 
- \frac{ r^{-2a-an+bn} }{ -r^b+r^{2a} }  + \frac{ r^{-2a+an} }{ -r^b+r^{2a} }  
\Biggr]
\end{split}
\end{equation}

\begin{equation}
\begin{split}
F^2_{mn}(a,b) = r^{a(n+m) + b} &\Biggl[
\frac{ r^{2a - an +b(n+m)} }{ 1-r^{2a+b} } + \frac{ r^{-b-an-bn} }{ 1-r^{2a+b} }\\
&+\frac{ r^{-an+b(n+m)} +r^{-b-an}-r^{-an}}{ r^b-1 }
 - \frac{ r^{-bn-b-an} }{ 1 - r^b }
\Biggr]
\end{split}
\end{equation}

\begin{equation}
\begin{split}
F^3_{mn}(a,b) = r^{a(n-m) + b} &\Biggl[
\frac{ r^{-an+2a+bn} }{ 1-r^{2a+b} } + \frac{ r^{-b+an+2am} }{ 1-r^{2a+b} }
+\frac{ r^{-an+bn} }{ r^b-1 }\\
& - \frac{ r^{-an+b(n+m)} }{ r^b-1 } 
- \frac{ r^{-an+b(n+m)} }{ -r^b+r^{2a} }  + \frac{ r^{an+2am} }{ -r^b+r^{2a} }  
\Biggr]
\end{split}
\end{equation}

\item{} In the zero temperature limit, $r\rightarrow 0$.
We then see that the free energy is
$1+20+12-1-32=0$, which is what is expected from supersymmetry.
\item{}We obtain parts of the free energy to three loop order
\item{}We check our three-loop computation by taking the zero
temperature limit of it and finding that the free energy vanishes
as the temperature is taken to zero.  This is expected as a result
of the supersymmetry of the model which is restored at zero
temperature.
\item{}We also find that the $\phi_n$-independent part of the free
energy vanishes to this order.  Note that this is not the same as
the zero temperature limit.
\item{}We find the full shift in $T_c$ to order $\lambda^2$.  It is
given in eqn. (\ref{tc}).
\item{}The form of the effective action that we find confirms
the first order nature of the phase transition at weak
coupling.

\end{itemize}

\section*{Acknowledgments}

This work was partially supported by NSERC of Canada, the String Theory
Collaborative Research Group of the Pacific Institute for Mathematical
Sciences and the Strings and Particles Collaborative Research Team of the
Pacific Institute for Theoretical Physics.

\section*{Appendix A: Three loops}

To do perturbation theory, we split the Euclidean action into
three parts, a free action, an interaction action with three-point
vertices and an interaction action with four-point vertices,

\begin{eqnarray}
S_0 =\int_{-\beta/2}^{\beta/2}d\tau
\Tr \left( \frac{1}{2}X^i\left(- D^2+(\mu/6)^2\right) X^i +
 \frac{1}{2}X^{\bar a}\left(- D^2+(\mu/3)^2\right) X^{\bar a} +
 \right. \nonumber \\ \left. +
\psi^{\dagger I \alpha}\left(- D+\mu/4\right) \psi_{I
\alpha}\right)
\end{eqnarray}

\begin{eqnarray}
S_3=\int_{-\beta/2}^{\beta/2}d\tau R^{\frac{3}{2}}\Tr\left( i\mu\frac{1}{3}
\epsilon_{\bar a \bar b \bar c} X^{\bar a} X^{\bar b} X^{\bar c}
  +  \psi^{\dagger I \alpha} \sigma^{\bar a}_\alpha
{}^\beta [X^{\bar a}, \psi_{I \beta}] -  ~~~~~~~
\right. \nonumber \\ \left. -\frac{1}{2}
\epsilon_{\alpha \beta} \psi^{\dagger \alpha I} {\sf g}^i_{IJ}
[X^i, \psi^{\dagger \beta J}] + \frac{1}{2} \epsilon^{\alpha
\beta}
\psi_{\alpha I} ({\sf g}^{i \dagger})^{IJ} [X^i, \psi_{\alpha J}]
\right)
\end{eqnarray}

\begin{equation}
S_4=-\int_{-\beta/2}^{\beta/2}d\tau \frac{R^3}{4}\Tr\left(  [X^i ,
X^j]^2+
[X^{\bar a}, X^{\bar b}]^2 +  2 [ X^{\bar a}, X^i]^2 \right)
\end{equation}
respectively.

The effective action $S_{\rm eff}[A]$ is obtained by a perturbative expansion of the
partition function.
\begin{equation}
e^{-S_{\rm eff}[A]}=\int
[dX][d\psi]e^{-S_0-S_3-S_4} ~~~,~~~ S_{\rm eff}[A]=1-<e^{-S_3-S_4}>
\end{equation}
where the bracket indicates the {\bf connected part} of the expectation value in the free theory.
Taylor expanding the exponential of interactions and retaining
those terms which have non-zero contributions to two loop order
gives
\begin{eqnarray}
S_{\rm
eff}=\sum_{a\neq b}\ln\left[\frac{  \det^{3/2}\left(
-D_{ab}^2+\frac{\mu^2}{9}\right) \det^{3} \left(
-D_{ab}^2+\frac{\mu^2}{36}\right)}{{\det}\left(-D_{ab}\right)
 \det^8\left(
-D_{ab}+\frac{\mu}{4}\right) }\right]+
~~~~~~~~~~~~~~~~~~~~~~~~~~~~~
\nonumber \\ + \left( <S_4>-\frac{1}{2}<S_3^2>\right)  - \left(
\frac{1}{2}<S_4^2>-\frac{1}{2}<S_3^2S_4>+\frac{1}{24}<S_3^4>\right)+\ldots
\end{eqnarray}
The first line on the right-hand-side is one-loop, the bracket
after it
contains two-loop and the next bracket the three-loop contributions.

The appropriate two-loop Feynman diagrams are as follows, where a
dotted line represents a fermion, and a solid line either flavour
of scalar.

\vspace*{0.4cm}
\[
\parbox{20mm}{
\begin{fmfgraph}(13,13)
\fmfleft{i}
\fmfright{o}
\fmf{phantom}{i,v}
\fmf{plain,tension=0.45}{v,v}
\fmf{phantom}{v,o}
\fmf{plain,left=90,tension=0.45}{v,v}
\fmfdot{v}
\end{fmfgraph}}\qquad
\parbox{20mm}{
\begin{fmfgraph}(20,20)
\fmfleft{i}
\fmfright{o}
\fmf{plain,left,tension=.2}{i,o}
\fmf{plain,tension=.2}{i,o}
\fmf{plain,right,tension=.2}{i,o}
\fmfdot{i,o}
\end{fmfgraph}} \qquad \qquad\parbox{20mm}{
\begin{fmfgraph}(20,20)
\fmfleft{i}
\fmfright{o}
\fmf{dashes,left,tension=.2}{i,o}
\fmf{plain,tension=.2}{i,o}
\fmf{dashes,right,tension=.2}{i,o}
\fmfdot{i,o}
\end{fmfgraph}}  
\]
\vspace*{0.4cm}

\noindent At three loops there are many more diagrams. The following figures
show the three loop diagrams and their associated zero temperature limits. 
The letter ``P" denotes the circulation of the $m=\mu/3$ scalar, while ``Q" 
denotes the $m=\mu/6$ scalar. Where there is no indication, it is obvious 
from the allowed interactions dictated by the action.

\subsubsection*{Cat's Eye Diagram}

\[
\parbox{20mm}{
\includegraphics*[bb= 215 483 385 627,width=1in, height=0.847in]{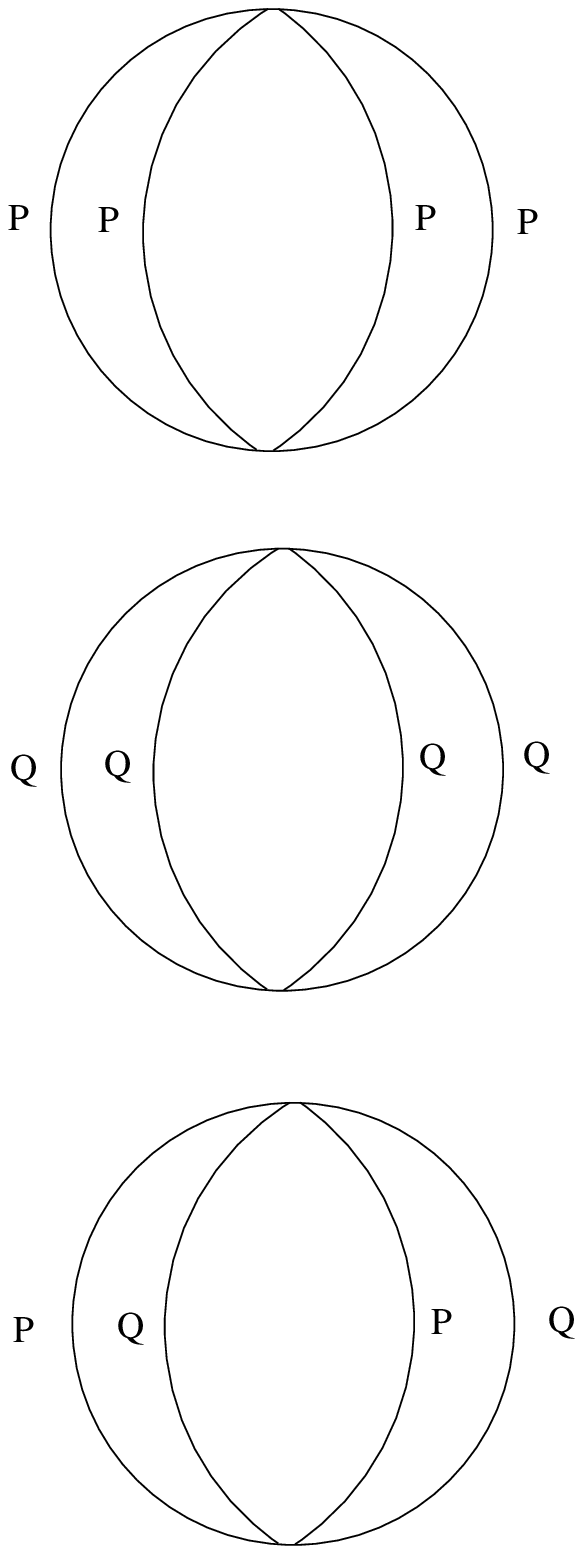}}
\qquad
=\frac{2187}{64} \frac{\beta R^6}{\mu^5}\quad \parbox{20mm}{
\includegraphics*[bb= 215 327 385 470,width=1in, height=0.847in]{catsi.ps}}
\qquad
=\frac{10935}{2} \frac{\beta R^6}{\mu^5}  \]

\[\parbox{20mm}{
\includegraphics*[bb= 215 168 395 310,width=1.074in, height=0.847in]{catsi.ps}}
\qquad
=\frac{2187}{2} \frac{\beta R^6}{\mu^5}
\]

\subsubsection*{Triple Bubble Diagram}

\begin{center}
\[
\parbox{20mm}{
\includegraphics*[bb= 180 530 418 616,width=1.31in, height=0.473in]{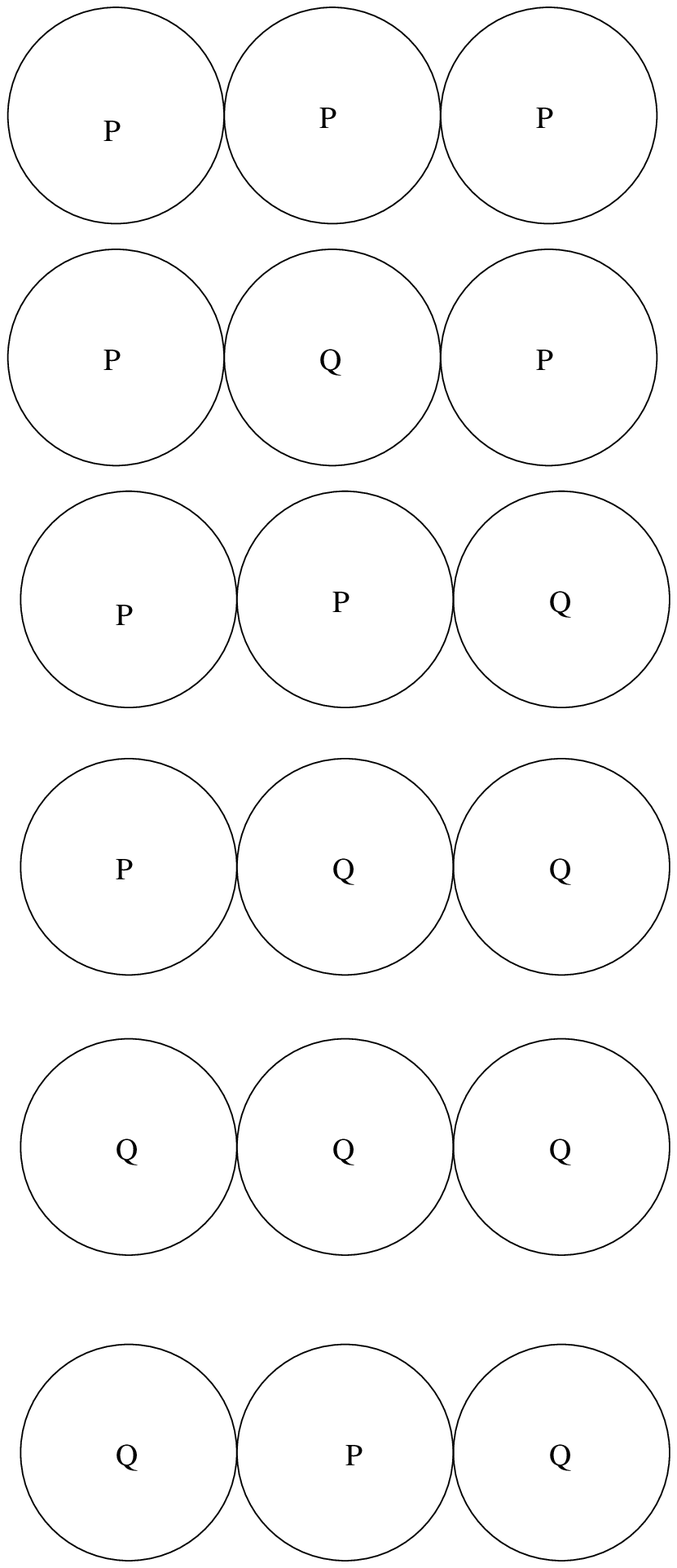}}
\qquad\qquad =6561\, \frac{\beta R^6}{\mu^5}\qquad \parbox{20mm}{
\includegraphics*[bb= 180 143 418 229,width=1.31in, height=0.473in]{trip_bubble.ps}}
\qquad\qquad =6561\, \frac{\beta R^6}{\mu^5}
\]
\[
\parbox{20mm}{
\includegraphics*[bb= 180 445 418 531,width=1.31in, height=0.473in]{trip_bubble.ps}}
\qquad\qquad =2187\, \frac{\beta R^6}{\mu^5}\qquad \parbox{20mm}{
\includegraphics*[bb= 180 350 418 436,width=1.31in, height=0.473in]{trip_bubble.ps}}
\qquad\qquad =43740\, \frac{\beta R^6}{\mu^5}
\]
\[
\parbox{20mm}{
\includegraphics*[bb= 180 615 418 701,width=1.31in, height=0.473in]{trip_bubble.ps}}
\qquad\qquad =\frac{729}{4} \frac{\beta R^6}{\mu^5}\qquad \parbox{20mm}{
\includegraphics*[bb= 180 250 418 336,width=1.31in, height=0.473in]{trip_bubble.ps}}
\qquad\qquad =72900\, \frac{\beta R^6}{\mu^5}\]
\end{center}

\subsubsection*{Theta-Bubble Diagram}

\begin{center}
\[
\parbox{20mm}{
\includegraphics*[bb= 310 385 460 490,width=0.726in, height=0.508in]{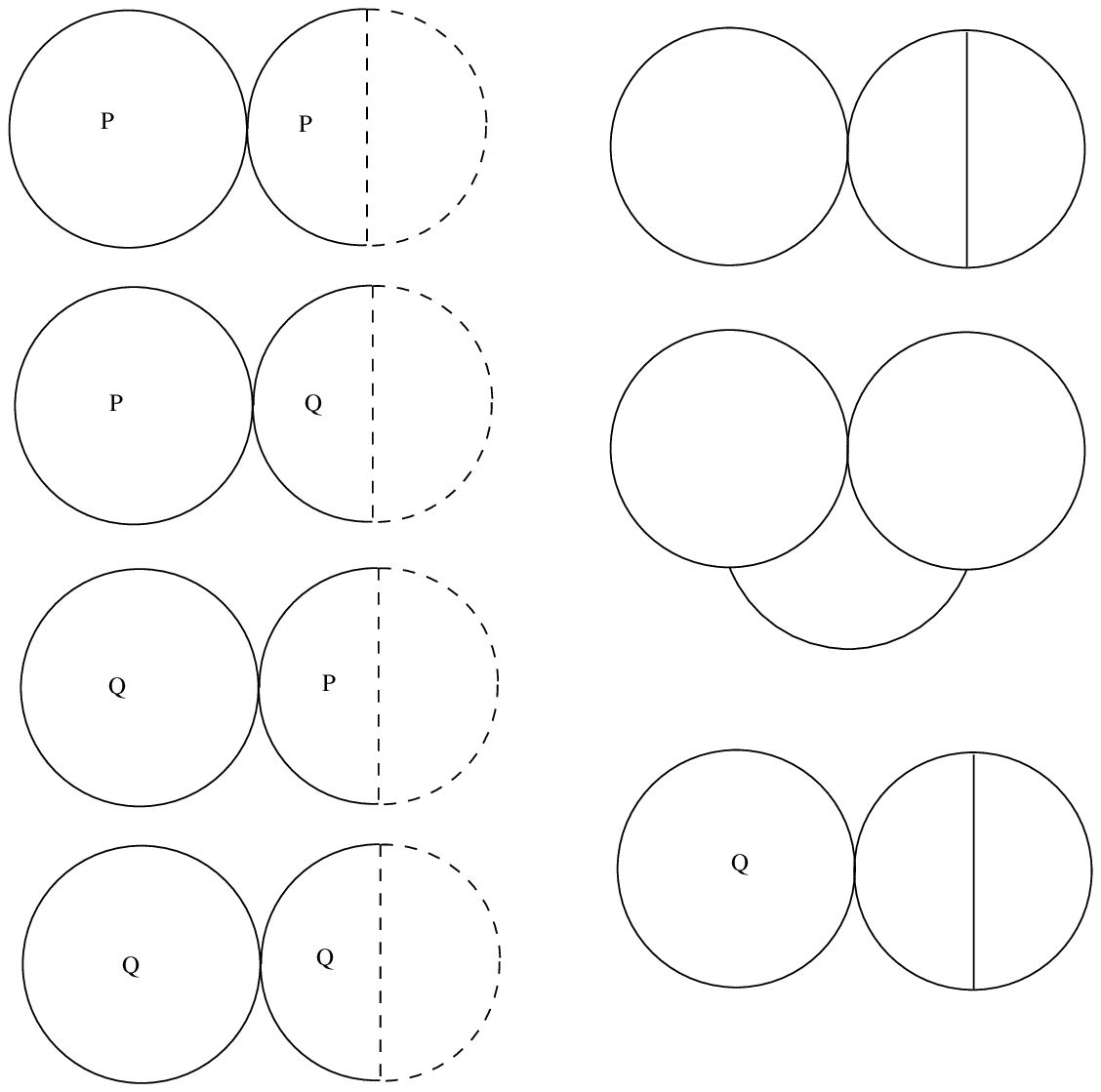}}
\qquad =-\frac{10935}{32} \frac{\beta R^6}{\mu^5}  \qquad \qquad \parbox{20mm}{
\includegraphics*[bb= 310 500 460 578,width=0.907in, height=0.473in]{th_bubble.ps}}
\qquad =-729\, \frac{\beta R^6}{\mu^5}  \]

\[
\parbox{20mm}{
\includegraphics*[bb= 132 505 282 583,width=0.907in, height=0.473in]{th_bubble.ps}}
\qquad \qquad \qquad =0 \qquad \qquad\parbox{20mm}{
\includegraphics*[bb= 310 287 460 365,width=0.907in, height=0.473in]{th_bubble.ps}}
\qquad  =-4374\, \frac{\beta R^6}{\mu^5}\]

\[
\parbox{20mm}{
\includegraphics*[bb= 133 424 283 502,width=0.907in, height=0.473in]{th_bubble.ps}}
\quad   =-43740\, \frac{\beta R^6}{\mu^5} \quad 
\parbox{20mm}{
\includegraphics*[bb= 136 341 286 419,width=0.907in, height=0.473in]{th_bubble.ps}}
\quad =0 \quad \parbox{20mm}{
\includegraphics*[bb= 136 260 286 338,width=0.907in, height=0.473in]{th_bubble.ps}}
\quad  =-145800\, \frac{\beta R^6}{\mu^5}\]

\end{center}

\subsubsection*{Circle-T Diagram}

\begin{center}
\[
\parbox{20mm}{
\includegraphics*[bb= 193 260 296 360,width=0.7in, height=0.688in]{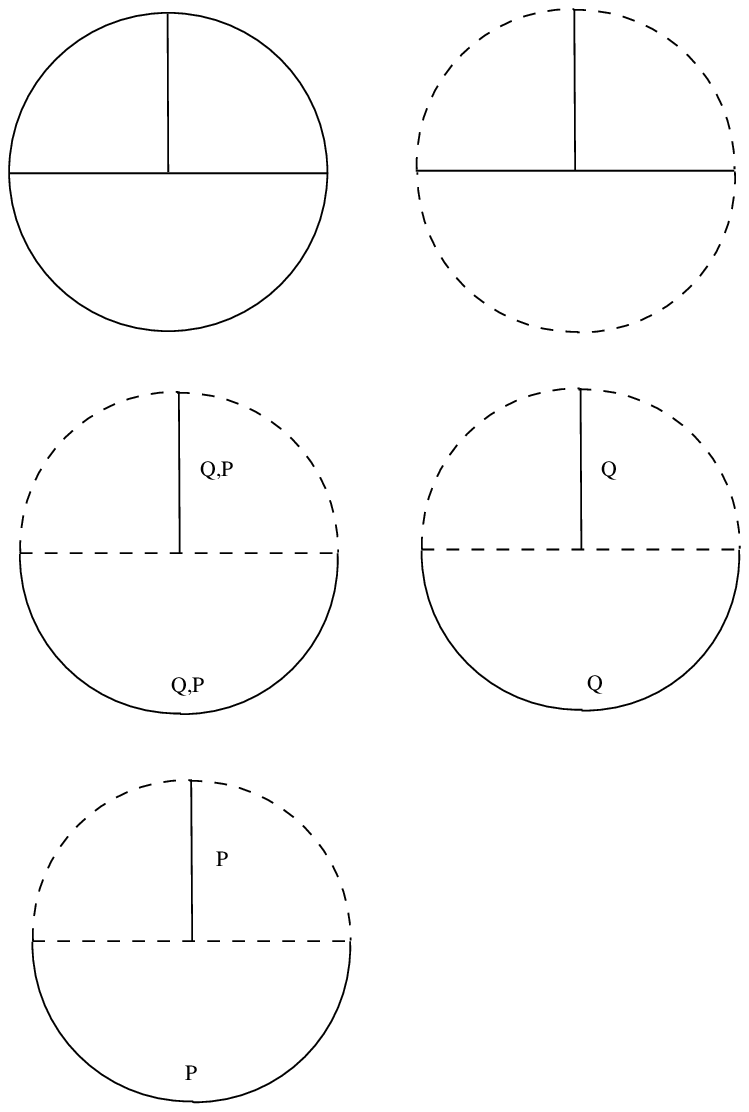}}
\qquad \qquad = 0 \qquad\qquad \parbox{20mm}{
\includegraphics*[bb= 190 372 293 472,width=0.7in, height=0.688in]{T.ps}}
\qquad =2916\, \frac{\beta R^6}{\mu^5} \]

\[
\parbox{20mm}{
\includegraphics*[bb= 304 372 407 472,width=0.7in, height=0.688in]{T.ps}}
\quad =7776\, \frac{\beta R^6}{\mu^5}\quad 
\parbox{20mm}{
\includegraphics*[bb= 304 482 407 582,width=0.7in, height=0.688in]{T.ps}}
\quad =0  \quad \parbox{20mm}{
\includegraphics*[bb= 185 482 288 582,width=0.7in, height=0.688in]{T.ps}}
\quad =\frac{6561}{64} \frac{\beta R^6}{\mu^5} \]

\end{center}

\subsubsection*{Two-Rung Ladder Diagrams}

%\begin{center}
\[
\parbox{20mm}{
\includegraphics*[bb= 164 300 244 420,width=0.6in, height=0.9in]{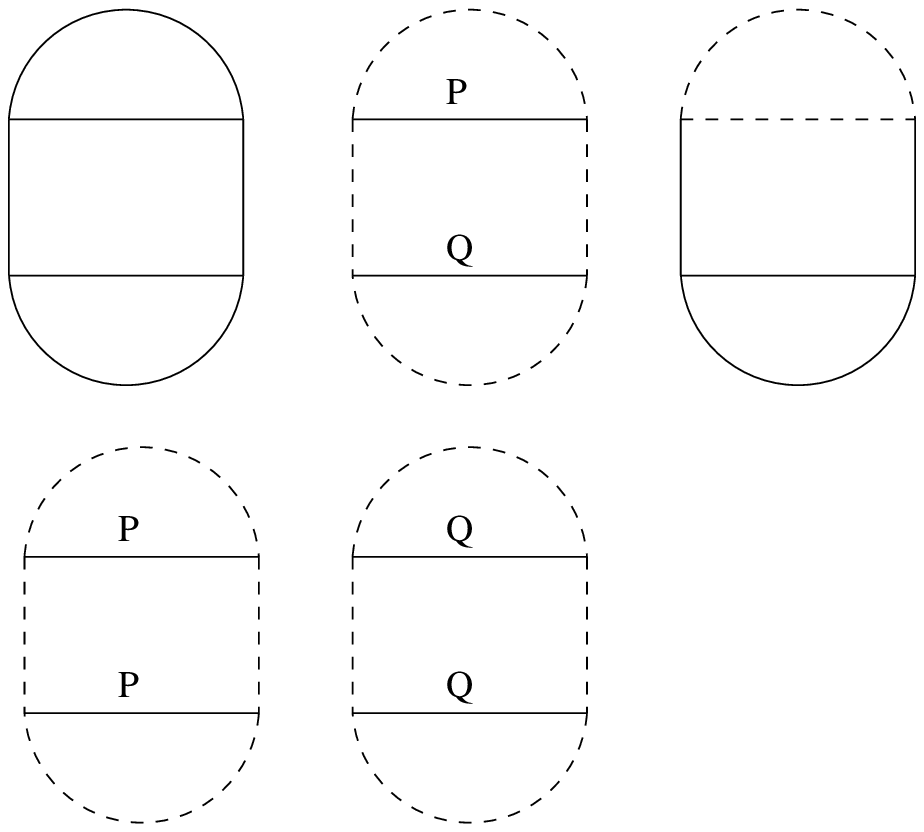}}
\qquad \qquad = 0 \qquad\qquad \parbox{20mm}{
\includegraphics*[bb= 256 300 337 420,width=0.6in, height=0.9in]{dblbar.ps}}
\qquad  =-34992\, \frac{\beta R^6}{\mu^5}\]

\[
\parbox{20mm}{
\includegraphics*[bb= 313 360 393 480,width=0.6in, height=0.9in]{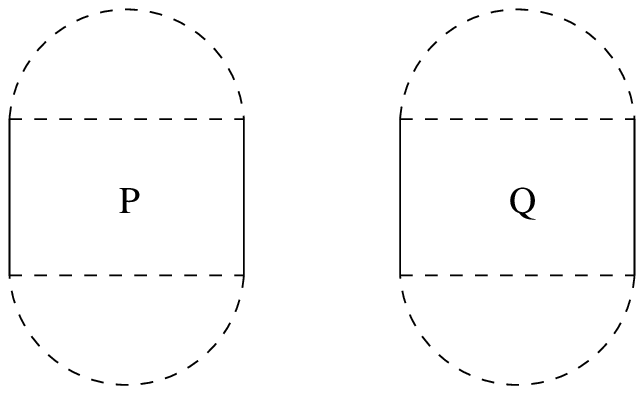}}
\qquad \quad =73872\, \frac{\beta R^6}{\mu^5} \qquad \qquad\parbox{20mm}{
\includegraphics*[bb= 200 360 280 480,width=0.6in, height=0.9in]{dblbar2.ps}}
\qquad \qquad =0\]

\[
\parbox{20mm}{
\includegraphics*[bb= 257 425 337 545,width=0.6in, height=0.9in]{dblbar.ps}}
\quad =5832\, \frac{\beta R^6}{\mu^5}\quad
\parbox{20mm}{
\includegraphics*[bb= 355 425 435 545,width=0.6in, height=0.9in]{dblbar.ps}}
\quad =0 \quad \parbox{20mm}{
\includegraphics*[bb= 160 425 240 545,width=0.6in, height=0.9in]{dblbar.ps}}
\quad =\frac{24057}{32} \frac{\beta R^6}{\mu^5}
\]

\noindent One can check that the sum of the above factors is zero, as garaunteed
by SUSY.

The vertices for these diagrams can be deduced from the
interaction terms in the action.  The propagators depend on
temperature and are discussed in the appendix below.

The technique that we use is to work in time, rather than momentum
representation of the propagators.   The integration over times is
elementary and most of the work involves extracting the moments in
$\eta_a\equiv e^{i\beta A_a}$.  Liberal use of  computer algebra
systems was used to
unravel the expansion of diagrams in this object.

\section*{Appendix B: Thermal Green functions}

The free field correlation function of the scalar field is
$$
\left< X^i_{ab}(\tau)X^j_{cd}(0)\right>=
\delta^{ij}\delta_{ad}\delta_{bc}\left[g_1(\tau)\right]_{ab} $$
where the Green function is
\begin{eqnarray}\label{g1}
\left[g_1(\tau)\right]_{ab}&=&(\tau|\frac{1}{-D_{ab}^2+
(\mu/6)^2}|0)\\&=&-\frac{\beta\eta^{-\tau/\beta}}{4\ln
r}\left[ \left(
\frac{1}{1-\frac{1}{\eta}r^2}\left[r^2\right]^{\tau/\beta}
+\frac{\eta r^2}{1-\eta
r^2}\left[r^2\right]^{-\tau/\beta}\right)\theta(\tau)+ \right.
\nonumber \\&& ~~~~~~~~~~~~~~~\left. +\left(
\frac{\frac{1}{\eta}r^2}{1-\frac{1}{\eta}r^2}\left[r^2\right]^{\tau/\beta}
+\frac{1}{1-\eta
r^2}\left[r^2\right]^{-\tau/\beta}\right)\theta(-\tau) \right]
\nonumber
\end{eqnarray} where $\eta=e^{i\beta A_{ab}}=\eta_a/\eta_b$ and
$r=e^{-\beta\mu/12}$.  This Green function is defined on the
interval $\tau/\beta\in(-1/2,1/2)$, with $g_1(1/2)=g_1(-1/2)$ and
it must be extended periodically to all real values
$\tau/\beta+$integers. An alternative expression which is
sometimes useful is
\begin{eqnarray}
g_1(\tau)= -\frac{\beta\eta^{-\tau/\beta}}{4\ln r}\left(
\left[r^2\right]^{\left|\tau\right|/\beta}+\sum_{n=1}^\infty
r^{2n}\left[ \eta^{-n}\left[r^2\right]^{\tau/\beta}
+\eta^n\left[r^2\right]^{-\tau/\beta}\right] \right)\nonumber
\end{eqnarray}
The first term is the zero temperature Euclidean green function
and the last two terms are the homogeneous solutions of the
Euclidean wave equation which must be added to the first term in
order to satisfy the periodic boundary conditions.  Note that both
green functions have a coefficient $\eta^{-\tau/\beta}$. This
factor cancels identically in all vacuum diagrams.  It has been
separated explicitly since some loop integrations are easier once
it is canceled.

Similarly,
$$
\left< X^{\bar a}_{ab}(\tau)X^{\bar b}_{cd}(0)\right>_0=
\delta^{\bar a\bar
b}\delta_{ad}\delta_{bc}\left[g_2(\tau)\right]_{ab}
$$
\begin{eqnarray}\label{g2}
\left[g_2(\tau)\right]_{ab}&=&(\tau|\frac{1}{-D_{ab}^2+(\mu/3)^2}|0)
\\&=&-\frac{\beta\eta^{-\tau/\beta}}{8\ln r}\left[ \left(
\frac{1}{1-\frac{1}{\eta}r^4}\left[r^4\right]^{\tau/\beta}
+\frac{\eta r^4}{1-\eta
r^4}\left[r^4\right]^{-\tau/\beta}\right)\theta(\tau)+ \right.
\nonumber \\ &&~~~~~~~~~~~~~~~~\left. +\left(
\frac{\frac{1}{\eta}r^4}{1-\frac{1}{\eta}r^4}\left[r^4\right]^{\tau/\beta}
+\frac{1}{1-\eta
r^4}\left[r^4\right]^{-\tau/\beta}\right)\theta(-\tau) \right]
\nonumber \\&&=-\frac{\beta\eta^{-\tau/\beta}}{8\ln r}\left(
\left[r^4\right]^{\left|\tau\right|/\beta}+\sum_{n=1}^\infty
r^{4n}\left[ \eta^{-n}\left[r^4\right]^{\tau/\beta}
+\eta^n\left[r^4\right]^{-\tau/\beta}\right] \right)\nonumber
\end{eqnarray}

The Fermion Green function is
\begin{equation}\label{gf}
\left[g_f(\tau)\right]_{ab}=(\tau|\frac{1}{-D_{ab}+(\mu/4)}|0)=  -
\eta^{-\tau/\beta}
\frac{\left[r^3\right]^{\tau/\beta}}{1+r^3/\eta}
\left(\theta(\tau)-\frac{1}{\eta}r^3\theta(-\tau)\right)
\end{equation}
It is antiperiodic $g_f(-1/2)=-g_f(1/2)$.

Note that the expressions in (\ref{g1}),(\ref{g2}) are unchanged
by the replacement $r\to 1/r$.

\end{fmffile}
\end{document}